\documentclass{article}
\usepackage{amsmath}
\usepackage[psamsfonts]{amssymb}
\usepackage{cmmib57}
\usepackage{diagrams}
\newcommand{\bPf}{\par\vspace*{-4pt}\indent{\sc Proof.}\enskip}
\newcommand{\ePf}{\medskip}
\def\QED{\hskip0.1em\hfill\null\ \null\nobreak\hfill\kern3pt\vbox{\hrule\hbox
    {\vrule\kern1pt\vbox{\kern1.7pt\hbox{$\scriptscriptstyle{QED}$}
     \kern0.2pt}\kern1pt\vrule}\hrule}}

\def\END{\hskip0.1em\hfill\null\ \null\nobreak\hfill\kern3pt\vbox{\hrule\hbox
    {\vrule\kern1pt\vbox{\kern1.7pt\hbox{$\,\,\,\vspace{5pt}$}
     \kern0.2pt}\kern1pt\vrule}\hrule}}
\newtheorem{theorem}{Theorem}
\newtheorem{lemma}{Lemma}
\newtheorem{corollary}{Corollary}
\newtheorem{proposition}{Proposition}
\newtheorem{remark}{Remark}
\newtheorem{definition}{Definition}
\newtheorem{example}{Example}
\newcommand{\bEq}{\begin{eqnarray}}
\newcommand{\eEq}{\end{eqnarray}}
\newcommand{\beq}{\begin{eqnarray*}}
\newcommand{\eeq}{\end{eqnarray*}}
\newcommand{\bDf}{\begin{definition}\em}
\newcommand{\eDf}{\end{definition}}
\newcommand{\bLm}{\begin{lemma}}
\newcommand{\eLm}{\end{lemma}}
\newcommand{\bPr}{\begin{proposition}}
\newcommand{\ePr}{\end{proposition}}
\newcommand{\bTh}{\begin{theorem}}
\newcommand{\eTh}{\end{theorem}}
\newcommand{\bCr}{\begin{corollary}}
\newcommand{\eCr}{\end{corollary}}
\newcommand{\bRm}{\begin{remark}\em}
\newcommand{\eRm}{\end{remark}}
\newcommand{\bEx}{\begin{example}\em}
\newcommand{\eEx}{\end{example}}

\newcommand{\ie}{{\em i.e$.$} }
\newcommand{\eg}{{\em e.g$.$} }
\newcommand{\R}{I\!\!R}

\newcommand{\mto}{\mapsto}

\newcommand{\der}{\partial}

\DeclareMathOperator{\byd}{{\raisebox{.1ex}{:}{=}}}

\newcommand{\ucar}[1]{\underset{#1}{\times}}

\newcommand{\owed}[1]{\overset{#1}{\wedge}}
\newcommand{\balp}{\boldsymbol{\alp}}
\newcommand{\bbet}{\boldsymbol{\bet}}

\newcommand{\blam}{\boldsymbol{\lam}}
\newcommand{\bmu}{\boldsymbol{\mu}}
\newcommand{\bnu}{\boldsymbol{\nu}}

\newcommand{\bsig}{\boldsymbol{\sig}}

\newcommand{\cA}{\mathcal{A}}
\newcommand{\cC}{\mathcal{C}}

\newcommand{\cE}{\mathcal{E}}

\newcommand{\cJ}{\mathcal{J}}

\newcommand{\cT}{\mathcal{T}}
\newcommand{\cZ}{\mathcal{Z}}
\newcommand{\cL}{\mathcal{L}}
\newcommand{\by}{\boldsymbol{y}}

\newcommand{\bF}{\boldsymbol{F}}
\newcommand{\bC}{\boldsymbol{C}}
\newcommand{\bG}{\boldsymbol{G}}

\newcommand{\bP}{\boldsymbol{P}}

\newcommand{\bX}{\boldsymbol{X}}
\newcommand{\bY}{\boldsymbol{Y}}
\newcommand{\bW}{\boldsymbol{W}}

\newcommand{\sub}{\subset}

\newcommand{\wed}{\wedge}
\newcommand{\com}{\!\circ\!}
\newcommand{\ten}{\!\otimes\!}
\newcommand{\alp}{\alpha}
\newcommand{\bet}{\beta}
\newcommand{\gam}{\gamma}
\newcommand{\del}{\delta}
\newcommand{\eps}{\epsilon}
\newcommand{\zet}{\zeta}

\newcommand{\lam}{\lambda}
\newcommand{\sig}{\sigma}

\newcommand{\ome}{\omega}
\newcommand{\Gam}{\Gamma}

\newcommand{\Lam}{\Lambda}

\newcommand{\vartht}{\vartheta}

\newcommand{\For}{{\Lambda}}
\newcommand{\Con}{{\mathcal{C}}}
\newcommand{\Hor}{{\mathcal{H}}}
\newcommand{\Var}{{\mathcal{V}}}
\newcommand{\Thd}{{\Theta}}
\title{\large{{\bf Covariant gauge-natural conservation laws}}}
\author{{\normalsize M.
Palese and E. Winterroth}\thanks{Both of them supported by
GNFM of INdAM and University of Torino.}
\\{\footnotesize Department of Mathematics,
University of Torino}
\\{\footnotesize via C. Alberto 10, 10123 Torino, Italy}\\ 
{\footnotesize e--mails: 
{\sc palese@dm.unito.it, ekkehart@dm.unito.it}}}
\date{}
\begin{document}

\maketitle

\begin{abstract}

When a gauge-natural invariant variational principle is
assigned, to determine {\em canonical} covariant conservation laws, 
the vertical part of
gauge-natural lifts of infinitesimal principal automorphisms -- 
defining infinitesimal
variations of sections of gauge-natural bundles -- must satisfy 
generalized Jacobi equations
for the gauge-natural invariant Lagrangian. {\em Vice versa} all vertical parts
of gauge-natural lifts of infinitesimal principal automorphisms which 
are in the kernel of
generalized Jacobi morphisms are generators of canonical covariant currents and
superpotentials. In
particular, only a few gauge-natural lifts can be considered as {\em 
canonical} generators of covariant
gauge-natural physical charges.
\medskip


\end{abstract}

%-------------------------------------------------------------------------%
\section{Introduction}
%-------------------------------------------------------------------------%

In
\cite{AnBe51,Vari1} the general program was started of
defining covariant conservation laws for field theories as generators 
of infinitesimal
transformations of the basis manifold. In
the classical Lagrangian formulation of field theories the 
description of symmetries and conserved quantities
amounts to define suitable (vector) densities  which generate the 
conserved currents; in all
relevant physical theories this densities are found to be the 
divergence of skew--symmetric
(tensor) densities, which are called {\em superpotentials} for the 
conserved currents. It
is also well known that the importance of superpotentials relies on 
the fact that they can
be integrated to provide conserved quantities associated with the 
conserved currents {\em
via} the Stokes theorem (see \eg \cite{FeFr91} and references quoted 
therein). Within such a procedure,
the generalized Bianchi identities for geometric field
theories - introduced by Bergman to get (after an integration by 
parts procedure) a consistent
equation between (local) divergences within the first variation formula -
are in fact necessary and (locally) sufficient conditions for the 
conserved current
$\epsilon$ to be not only closed but also the divergence of a 
skew-symmetric (tensor) density (a superpotential)
along solutions of the  Euler--Lagrange equations.
However, to ``covariantize'' such a derivation of Bianchi
identities and superpotentials, background metrics or
(fibered) connections have to be fixed {\em a priori} (see \eg
\cite{Vari2,FFP01,FeFr91} and the wide literature quoted therein).
The outcoming of such an {\em ad hoc} procedure is, notably, a 
triviality result concerning
{\em existence of global superpotentials} for gauge-natural field 
theories despite of the topology of the basis of the gauge-natural 
bundle (see \eg the review in \cite{Mat03}). 

We shall show that, when a gauge-natural invariant variational principle is
assigned, to determine {\em canonical} (\ie completely determined by 
the variational problem and its invariance properties) covariant 
conservation laws, the vertical part of
gauge-natural lifts of infinitesimal principal automorphisms must 
satisfy generalized Jacobi equations
for the gauge-natural invariant Lagrangian. {\em Vice versa} all vertical parts
of gauge-natural lifts of infinitesimal principal automorphisms which 
are in the kernel of the
generalized Jacobi morphism are generators of canonical covariant currents and
superpotentials. 

This result is the outcoming of the following 
facts:
\begin{enumerate}
\item the role played by the invariance of a 
given variational problem of order $s$ on a fibered manifold $\bY \to 
\bX$  with respect to the contact structure induced by the affine 
fiberings $\pi^{s+1}_{s}: J_{s+1}\bY \to J_{s}\bY$ and its encoding 
through the Krupka's {\em finite order  variational sequence} 
language. It is fundamental to stress that such an invariance has to 
reflect too in a precise way on the nature of the conserved Noether 
currents associated with a given gauge-natural invariant Lagrangian 
(see Remark \ref{fundRem});
\item the Noether Theorems ({\em both of 
them}) take a quite particular form in the case of gauge-natural 
Lagrangian field theories (see \eg  \cite{FFP01,Mat03}) due to the 
fact that the generalized Lie derivative of sections of the 
gauge-natural bundles has special linearity properties and it is 
related with the vertical part (with respect to the splitting induced 
by the contact structure) of gauge-natural lifts of  right-invariant 
(also called principal) infinitesimal automorphisms of the underlying 
principal bundle (structure bundle);
\item the {\em second variation} 
of the action functional can be conveniently represented in the 
finite order variational sequence framework in terms of iterated {\em variational 
Lie derivatives} (the quotient Lie derivative of {\em variational 
morphisms} for  first introduced in \cite{FPV98a}) with respect to 
vertical parts  of gauge-natural lifts of principal infinitesimal 
automorphisms. In particular, {\em by resorting to the Second Noether 
Theorem}, in \cite{FrPa00,FrPa01} the second variation has been 
related with the generalized Jacobi morphism and in \cite{PaWi03} the 
relation of the kernel of generalized {\em gauge-natural} Jacobi 
morphism with the kernel of a fundamental morphism, the {\em 
(Bergman-)Bianchi} morphism, has been explicitly clarified in order 
to characterize Bianchi identities for geometric field theories in 
terms of a special class of gauge-natural lifts of infinitesimal 
principal automorphisms, namely those which have their vertical part 
in the kernel of the generalized gauge-natural Jacobi 
morphism.
\end{enumerate}

Here we claim that the indeterminacy 
appearing in the derivation of gauge-natural conserved charges (see 
the interesting papers \cite{GoMa03,Mat03}) - \ie the difficulty of 
relating in a natural way  infinitesimal gauge transfomations with 
infinitesimal transformations of the basis manifold (\eg of 
space-time) - can be solved by requiring the second variation to be 
zero too. Historically Jacobi equations were related to
the so--called {\em accessory problem} (see, {\em e.g.}
\cite{Car82,Run66}), where they are directly obtained as
the variation of the Euler--Lagrange equations
of a given Lagrangian. Thus they can be characterized {\em via} the 
Second Noether 
Theorem.

%---------------------------------------------------------------------------------------
\section{Variational sequences on gauge-natural bundles}
%---------------------------------------------------------------------------------------

Our framework is a fibered manifold $\pi : \bY \to \bX$,
with $\dim \bX = n$ and $\dim \bY = n+m$.
For $s \geq q \geq 0$ integers we are concerned with the $s$--jet 
space $J_s\bY$ of $s$--jet prolongations of
(local) sections
of $\pi$ (see \eg \cite{KMS93,MaMo83a,Sau89}); in particular, we set 
$J_0\bY \equiv \bY$. We recall the natural
fiberings
$\pi^s_q: J_s\bY \to J_q\bY$, $s \geq q$, $\pi^s: J_s\bY \to \bX$, and,
among these, the {\em affine\/} fiberings $\pi^{s}_{s-1}$.
We denote with $V\bY$ the vector subbundle of the tangent
bundle $T\bY$ of vectors on $\bY$ which are vertical with respect
to the fibering $\pi$.

Charts on $\bY$ adapted to $\pi$ are denoted by $(x^\sig ,y^i)$.  Greek
indices $\sig ,\mu ,\dots$ run from $1$ to $n$ and they label basis
coordinates, while
Latin indices $i,j,\dots$ run from $1$ to $m$ and label fibre coordinates,
unless otherwise specified.
We denote multi--indices of dimension $n$ by boldface Greek letters such as
$\balp = (\alp_1, \dots, \alp_n)$, with $0 \leq \alp_\mu$,
$\mu=1,\ldots,n$; by an abuse
of notation, we denote with $\sig$ the multi--index such that
$\alp_{\mu}=0$, if $\mu\neq \sig$, $\alp_{\mu}= 1$, if
$\mu=\sig$.
We also set $|\balp| \byd \alp_{1} + \dots + \alp_{n}$ and $\balp ! \byd
\alp_{1}! \dots \alp_{n}!$.
The charts induced on $J_s\bY$ are denoted by $(x^\sig,y^i_{\balp})$, with $0
\leq |\balp| \leq s$; in particular, we set $y^i_{\bf{0}}
\equiv y^i$. The local vector fields and forms of $J_s\bY$ induced by
the above coordinates are denoted by $(\der^{\balp}_i)$ and $(d^i_{\balp})$,
respectively.

In the theory of variational sequences a fundamental role is played by the
{\em contact maps\/} on jet spaces (see \cite{Kru90,MaMo83a,Vit98}).
Namely, for $s\geq 1$, we consider the natural complementary fibered
morphisms over $J_s\bY \to J_{s-1}\bY$
\beq
\mathcal{D} : J_s\bY \ucar{\bX} T\bX \to TJ_{s-1}\bY \,,
\qquad
\vartht : J_{s}\bY \ucar{J_{s-1}\bY} TJ_{s-1}\bY \to VJ_{s-1}\bY \,,
\eeq
with coordinate expressions, for $0 \leq |\balp| \leq s-1$, given by
\beq
\mathcal{D} &= d^\lam\ten {\mathcal{D}}_\lam = d^\lam\ten
(\der_\lam + y^j_{\balp+\lam}\der_j^{\balp}) \,,
\vartht &= \vartht^j_{\balp}\ten\der_j^{\balp} =
(d^j_{\balp}-y^j_{{\balp}+\lam}d^\lam)
\ten\der_j^{\balp} \,.
\eeq
which induce the following natural splitting:
\bEq
\label{jet connection}
J_{s}\bY\ucar{J_{s-1}\bY}T^*J_{s-1}\bY =\left(
J_s\bY\ucar{J_{s-1}\bY}T^*\bX\right) \oplus\cC^{*}_{s-1}[\bY]\,,
\eEq
where $\cC^{*}_{s-1}[\bY] \simeq J_{s}\bY \ucar{J_{s-1}\bY} V^{*}J_{s-1}\bY$.

If $f: J_{s}\bY \to \R$ is a function, then we set
$D_{\sig}f$ $\byd \mathcal{D}_{\sig} f$,
$D_{\balp+\sig}f$ $\byd D_{\sig} D_{\balp}f$, where $D_{\sig}$ is
the standard {\em formal derivative}.
Given a vector field $\Xi : J_{s}\bY \to TJ_{s}\bY$, the splitting
\eqref{jet connection} yields $\Xi \, \com \, \pi^{s+1}_{s} = \Xi_{H} 
+ \Xi_{V}$
where, if $\Xi = \Xi^{\gam}\der_{\gam} + \Xi^i_{\balp}\der^{\balp}_i$, then we
have $\Xi_{H} = \Xi^{\gam}D_{\gam}$ and
$\Xi_{V} = (\Xi^i_{\balp} - y^i_{\balp + \gam}\Xi^{\gam})
\der^{\balp}_{i}$. We shall call $\Xi_{H}$ and $\Xi_{V}$ the
horizontal and the vertical part of $\Xi$, respectively.

The splitting
\eqref{jet connection} induces also a decomposition of the
exterior differential on $\bY$,
$(\pi^{s}_{s-1})^*\com \,d = d_H + d_V$, where $d_H$ and $d_V$
are defined to be the {\em horizontal\/} and {\em vertical differential\/}.
The action of $d_H$ and $d_V$ on functions and $1$--forms
on $J_s\bY$ uniquely characterizes $d_H$ and $d_V$ (see, {\em e.g.},
\cite{Sau89,Vit98} for more details).
A {\em projectable vector field\/} on $\bY$ is defined to be a pair
$(\Xi,\xi)$, where $\Xi:\bY \to T\bY$ and $\xi: \bX \to T\bX$
are vector fields and $\Xi$ is a fibered morphism over $\xi$.
If there is no danger of confusion, we will denote simply by $\Xi$ a
projectable vector field $(\Xi,\xi)$.
A projectable vector field $(\Xi,\xi)$
can be prolonged by the flow functor to a projectable vector field
$(j_{s}\Xi, \xi)$, the coordinate expression of which can be found \eg in
\cite{Kru90,MaMo83a,Sau89,Vit98};
in particular, we have the following expressions
$(j_{s}\Xi)_{H} = \xi^{\sig} \, D_{\sig}$,
$(j_{s}\Xi)_{V} = D_{\balp}(\Xi_{V})^i \, \der_i^{\balp}$,
with $(\Xi_{V})^i = \xi^i - \, y^i_{\sig}\xi^{\sig}$, for the
horizontal and the vertical part of $j_{s}\Xi$, respectively.
 From now on, by an abuse of notation, we will write simply $j_{s}\Xi_{H}$ and
$j_{s}\Xi_{V}$.
In particular, we stress that $j_{s}\Xi_{V}$ can be seen as a fibered 
morphism: $j_{s}\Xi_{V}: J_{s+1}\bY\ucar{J_{s}\bY}J_{s}\bY\to 
J_{s+1}\bY\ucar{J_{s}\bY}J_{s}V\bY$.

%---------------------------------------------------------------
\subsection{Gauge-natural bundles}
%---------------------------------------------------------------

In the following, we shall develop a suitable geometrical setting 
which enables us to
define and investigate the fundamental concept of
conserved quantity in gauge-natural Lagrangian field theories.

An important generalization of natural field theories
\cite{Tra67} to gauge fields theories passed through
the concept of jet prolongation of a principal bundle and the
introduction of a very important geometric construction, namely the 
gauge-natural bundle
functor \cite{Ec81,KMS93}.

Let $\bP\to\bX$ be a principal bundle with structure group $\bG$.
Let $r\leq k$ be integers and $\bW^{(r,k)}\bP$ $\byd$ 
$J_{r}\bP\ucar{\bX}L_{k}(\bX)$,
where $L_{k}(\bX)$ is the bundle of $k$--frames
in $\bX$ \cite{Ec81,KMS93}, $\bW^{(r,k)}\bG \byd J_{r}\bG\odot GL_{k}(n)$
the semidirect product with respect to the action of $GL_{k}(n)$
on $J_{r}\bG$ given by the
jet composition and $GL_{k}(n)$ is the group of $k$--frames
in $\R^{n}$. Here we denote by $J_{r}\bG$ the space of 
$(r,n)$-velocities on $\bG$.
The bundle $\bW^{(r,k)}\bP$ is a principal bundle over $\bX$ with 
structure group
$\bW^{(r,k)}\bG$.
The right action of $\bW^{(r,k)}\bG$ on the fibers of $\bW^{(r,k)}\bP$
is defined by the composition of jets (see, {\em e.g.},
\cite{KMS93}).

\bDf
The principal bundle $\bW^{(r,k)}\bP$ (resp. the Lie group $\bW^{(r,k)}\bG$)
is said to be the {\em gauge-natural prolongation of order $(r,k)$
of $\bP$ (resp. of $\bG$)}.
\END\eDf

\bDf
We define the {\em vector}
bundle over $\bX$ of right--invariant infinitesimal automorphisms of $\bP$
by setting $\cA = T\bP/\bG$.

For $r\leq k$ we also define the {\em vector} bundle  over $\bX$ of 
right invariant
infinitesimal automorphisms of $\bW^{(r,k)}\bP$ by setting
$\cA^{(r,k)} \byd T\bW^{(r,k)}\bP/\bW^{(r,k)}\bG$.
\END\eDf

Let $\bF$ be any manifold and $\zet:\bW^{(r,k)}\bG\ucar{}\bF\to\bF$ be
a left action of $\bW^{(r,k)}\bG$ on $\bF$. There is a naturally defined
right action of $\bW^{(r,k)}\bG$ on $\bW^{(r,k)}\bP \times \bF$ so that
  we can associate in a standard way
to $\bW^{(r,k)}\bP$ the bundle, on the given basis $\bX$,
$\bY_{\zet} \byd \bW^{(r,k)}\bP\times_{\zet}\bF$ \cite{Ec81,KMS93}.

\bDf
We say $(\bY_{\zet},\bX,\pi_{\zet};\bF,\bG)$ to be the
{\em gauge-natural bundle} of order
$(r,k)$ associated to the principal bundle $\bW^{(r,k)}\bP$
by means of the left action $\zet$ of the group
$\bW^{(r,k)}\bG$ on the manifold $\bF$.
\END\eDf

\bRm
A principal automorphism $\Phi$ of $\bW^{(r,k)}\bP$ induces an
automorphism of the gauge-natural bundle by:
\bEq
\Phi_{\zet}:\bY_{\zet}\to\bY_{\zet}: [(j^{x}_{r}\gam,j^{0}_{k}t),
\hat{f}]_{\zet}\mto [\Phi(j^{x}_{r}\gam,j^{0}_{k}t),
\hat{f}]_{\zet}\,,
\eEq
where $\hat{f}\in \bF$ and $[\cdot, \cdot]_{\zet}$ is the equivalence class
induced by the action $\zet$.
\END\eRm

Denote by $\cT_{\bX}$ and $\cA^{(r,k)}$ the sheaf of
vector fields on $\bX$ and the sheaf of right invariant vector fields
on $\bW^{(r,k)}\bP$, respectively. A functorial mapping 
$\mathfrak{G}$ is defined
which lifts any right--invariant local automorphism $(\Phi,\phi)$ of the
principal bundle $W^{(r,k)}\bP$ into a unique local automorphism
$(\Phi_{\zet},\phi)$ of the associated bundle $\bY_{\zet}$.
Its infinitesimal version associates to any $\bar{\Xi} \in \cA^{(r,k)}$,
projectable over $\xi \in \cT_{\bX}$, a unique {\em projectable} vector field
$\hat{\Xi} \byd \mathfrak{G}(\bar{\Xi})$ on $\bY_{\zet}$ in the
following way:
\bEq
\mathfrak{G} : \bY_{\zet} \ucar{\bX} \cA^{(r,k)} \to T\bY_{\zet} \,:
(\by,\bar{\Xi}) \mto \hat{\Xi} (\by) \,,
\eEq
where, for any $\by \in \bY_{\zet}$, one sets: $\hat{\Xi}(\by)=
\frac{d}{dt} [(\Phi_{\zet \,t})(\by)]_{t=0}$,
and $\Phi_{\zet \,t}$ denotes the (local) flow corresponding to the
gauge-natural lift of $\Phi_{t}$.

This mapping fulfils the following properties:
\begin{enumerate}
\item {\em $\mathfrak{G}$ is linear over $id_{\bY_{\zet}}$};
\item we have $T\pi_{\zet}\circ\mathfrak{G} = id_{T\bX}\circ
\bar{\pi}^{(r,k)}$,
where $\bar{\pi}^{(r,k)}$ is the natural projection
$\bY_{\zet}\ucar{\bX}
\cA^{(r,k)} \to T\bX$;
\item $\mathfrak{G}$ is a homomorphism of Lie algebras: for any pair 
$(\bar{\Lam},\bar{\Xi})$ of vector fields in
$\cA^{(r,k)}$, we have
$
\mathfrak{G}([\bar{\Lam},\bar{\Xi}]) = [\mathfrak{G}(\bar{\Lam}), 
\mathfrak{G}(\bar{\Xi})]$;
\item in coordinates $
\mathfrak{G} = d^\mu \ten \der_\mu + d^{A}_{\bnu}
\ten (\cZ^{i\bnu}_{A} \der_{i}) + d^{\nu}_{\blam}
\ten (\cZ^{i\blam}_{\nu} \der_{i})$,
with $0<|\bnu|<k$, $1<|\blam|<r$ and
$\cZ^{i\bnu}_{A}$, $\cZ^{i\blam}_{\nu}$ $\in C^{\infty}(\bY_{\zet})$
are suitable functions which depend on the bundle, precisely on the 
fibers (see \cite{KMS93}).
\end{enumerate}

\bDf
The map $\mathfrak{G}$ is called the {\em gauge-natural lifting
functor}.
The  {\em projectable} vector field $(\hat{\Xi},\xi)\equiv 
\mathfrak{G}((\bar{\Xi},\xi))$ is
called the {\em
gauge-natural lift} of $(\bar{\Xi},\xi)$ to the bundle $\bY_{\zet}$.\END
\eDf

We shall consider  {\em variation vector fields} which are vertical 
parts of gauge-natural lifts of infinitesimal principal 
automorphisms. We recall that, due to the very definition of 
generalized Lie derivative of sections of gauge-natural bundles, 
variation vector fields are in fact formal Lie derivatives of 
sections with respect to gauge-natural lifts (see the item $4$ in the 
following). This will enables us to realize morphisms such as the 
Jacobi or the Bianchi morphisms in a very suitable way for our 
purposes.

\bDf({\bf Lie derivative of sections.})
Let $\gam$ be a (local) 
section of the gauge-natural bundle $\bY_{\zet}$, $\bar{\Xi}$
$\in \cA^{(r,k)}$ and $\hat\Xi$ its gauge-natural lift.
Following \cite{KMS93} we
define the {\em
generalized Lie derivative} of $\gam$ along the vector field
$\hat{\Xi}$ to be the (local) section $\pounds_{\bar{\Xi}} \gam : \bX 
\to V\bY_{\zet}$,
by setting:
$\pounds_{\bar{\Xi}} \gam = T\gam \circ \xi - \hat{\Xi} \circ \gam$.
\END\eDf

\bRm
This section is a vertical prolongation of $\gam$, \ie it satisfies
the property: $\nu_{\bY_{\zet}} \circ \pounds_{\Xi} \gam = \gam$,
where $\nu_{\bY_{\zet}}$ is the projection
$\nu_{\bY_{\zet}}: V\bY_{\zet} \to \bY_{\zet}$.
Its coordinate expression is given by
$(\pounds_{\bar{\Xi}}\gam)^{i} = \xi^{\sig} \der_{\sig} \gam ^{i} -
\hat\Xi^{i}(\gam)$. As customary we denote it by $\pounds_{\bar{\Xi}} 
\gam$ and not by $\pounds_{\hat{\Xi}} \gam$ because of the functorial 
correspondence between $\bar{\Xi}$ and $\hat{\Xi}$.
\eRm

\bRm\label{lie}
The Lie derivative operator acting on sections of gauge-natural
bundles satisfies the following
properties:
\begin{enumerate}\label{lie properties}
\item for any vector field $\bar{\Xi} \in \cA^{(r,k)}$, the
mapping $\gam \mto \pounds_{\bar{\Xi}}\gam$
is a first--order quasilinear differential operator;
\item for any local section $\gam$ of $\bY_{\zet}$, the mapping
$\bar{\Xi} \mto \pounds_{\bar{\Xi}}\gam$
is a linear differential operator;
\item we can regard $\pounds_{\bar{\Xi}}: J_{1}\bY_{\zet} \to V\bY_{\zet}$
as a morphism over the
basis $\bX$. In this case it is meaningful to consider the (standard)
jet
prolongation of $\pounds_{\bar{\Xi}}$, denoted by
$j_{s}\pounds_{\bar{\Xi}}:
J_{s+1}\bY_{\zet} \to VJ_{s}\bY_{\zet}$.
By using the canonical
isomorphism $VJ_{s}\bY_{\zet}\simeq J_{s}V\bY_{\zet}$, we have
$\pounds_{\bar{\Xi}}(j_{s}\gam) = j_{s} (\pounds_{\bar{\Xi}} \gam)$,
for any (local) section $\gam$ of $\bY_{\zet}$ and for any (local)
vector field $\bar{\Xi}\in \cA^{(r,k)}$.

\item as a consequence of linearity properties of gauge-natural 
lifts, we have 
$j_{s}\hat{\Xi}_{V}(\gam)=-\pounds_{j_{s}\bar{\Xi}}\gam$. In 
particular,  we can consider the Lie derivative of sections, 
$\pounds$, as a bundle morphism \cite{PaWi03}:
\beq
\pounds: J_{s+1}(\bY_{\zet} \ucar{\bX}
\cA^{(r,k)}) \to J_{s+1}\bY_{\zet}\ucar{J_{s}\bY_{\zet}}VJ_{s}\bY_{\zet}\,.
\eeq

\end{enumerate}
\END\eRm

%--------------------------------------------------------------------------%
\subsection{Variational Lie derivative of variational morphisms}
%--------------------------------------------------------------------------%

For the sake of simplifying notation, sometimes, we will omit the 
subscript $\zet$, so
that all our considerations shall refer to $\bY$ as a gauge-natural
bundle as defined above.

We shall be here concerned with some distinguished sheaves of forms on jet
spaces \cite{Kru90,Sau89,Vit98}. We shall in particular follow 
notation given in \cite{Vit98} to which the
reader is referred for details.
  For $s \geq 0$, we consider the standard sheaves $\For^{p}_{s}$
of $p$--forms on $J_s\bY$.
  For $0 \leq q \leq s $, we consider the sheaves $\Hor^{p}_{(s,q)}$ and
$\Hor^{p}_{s}$ of {\em horizontal forms} with respect to the 
projections $\pi^s_q$ and $\pi^s_0$, respectively.
  For $0 \leq q < s$, we consider the subsheaves $\Con^{p}_{(s,q)}
\sub \Hor^{p}_{(s,q)}$ and $\Con^{p}{_s} \sub
\Con^{p}_{(s+1,s)}$ of {\em contact forms}, \ie horizontal forms 
valued into $\cC^{*}_{s}[\bY]$(they have the property of vanishing 
along any section of the gauge-natural bundle).

According to \cite{Kru90,Vit98}, the fibered splitting
\eqref{jet connection} yields the {\em sheaf splitting}
$\Hor^{p}_{(s+1,s)}$ $=$ $\bigoplus_{t=0}^p$
$\Con^{p-t}_{(s+1,s)}$ $\wed\Hor^{t}_{s+1}$, which restricts to the inclusion
$\For^{p}_s$ $\sub$ $\bigoplus_{t=0}^{p}$
$\Con^{p-t}{_s}\wed\Hor^{t,}{_{s+1}^{h}}$,
where $\Hor^{p,}{_{s+1}^{h}}$ $\byd$ $h(\For^{p}_s)$ for $0 < p\leq 
n$ and the surjective map
$h$ is defined to be the restriction to $\For^{p}_{s}$ of the projection of
the above splitting onto the non--trivial summand with the highest
value of $t$. By an abuse of notation, let us denote by $d\ker h$ the sheaf
generated by the presheaf $d\ker h$ in the standard way.
We set $\Thd^{*}_{s}$ $\byd$ $\ker h$ $+$
$d\ker h$.

The $s$--th order quotient
{\em variational sequence} associated with the fibered manifold $\bY\to\bX$:
\beq
\diagramstyle[size=1.3em]
\begin{diagram}
0 & \rTo & \R_{Y} & \rTo & \For^{0}_s & \rTo^{\cE_{0}} &
\For^{1}_s/\Thd^{1}_s & \rTo^{\cE_{1}} & \For^{2}_s/\Thd^{2}_s & 
\rTo^{\cE_{2}} &
\dots & \rTo^{\cE_{I-1}} & \For^{I}_s/\Thd^{I}_s & \rTo^{\cE_{I}} &
\For^{I+1}_s & \rTo^{d} & 0\,,
\end{diagram}
\eeq
has been introduced by Krupka \cite{Kru90}.

To the aim of characterizing some fundamental morphisms for the 
calculus of variations  as sections of quotient sheaves and as 
corresponding differential quotient morphisms, let us consider the 
truncated variational sequence:
\beq
\diagramstyle[size=1.3em]
\begin{diagram}
0 &\rTo & \R_{Y} &\rTo & \Var^{0}_s & \rTo^{\cE_0} &
\Var^{1}_{s} & \rTo^{\cE_{1}} & \dots  & \rTo^{\cE_{n}} &
\Var^{n+1}_{s}  & \rTo^{\cE_{n+1}} & \cE_{n+1}(\Var^{n+1}_{s})
& \rTo^{\cE_{n+2}} &
0 \,,
\end{diagram}
\eeq
where, following \cite{Vit98}, the sheaves $\Var^{p}_{s}\byd
\Con^{p-n}_{s}\wed\Hor^{n,}{_{s+1}^h}/h(d\ker h)$ with $0\leq p\leq n+2$ are
suitable representations of the corresponding quotient
sheaves in the variational sequence by means of sheaves of sections of tensor
bundles.

Let $\alp\in\Con^{1}_s\wed\Hor^{n,}{_{s+1}^h}
\sub \Var^{n+1}_{s+1}$. Then there is a unique pair of
sheaf morphisms (\cite{Kol83,KoVi03,Vit98})
\bEq\label{first variation}
E_{\alp} \in \Con^{1}_{(2s,0)}\wed\Hor^{n,}{_{2s+1}^{h}} \,,
\qquad
F_{\alp} \in \Con^{1}_{(2s,s)} \wed \Hor^{n,}{_{2s+1}^h} \,,
\eEq
such that
$(\pi^{2s+1}_{s+1})^*\alp=E_{\alp}-F_{\alp}$,
and $F_\alp$ is {\em locally} of the form $F_{\alp} = d_{H}p_{\alp}$, 
with $p_{\alp}
\in \Con^{1}_{(2s-1,s-1)}\wed\Hor^{n-1}{_{2s}}$.

\bDf
Let $\gam \in \For^{n+1}_{s}$.
The morphism $E_{h(\gam)}\in\Var^{n+1}_{s}$ is called the
{\em generalized Euler--Lagrange morphism} associated with
$\gam$.
\END
\eDf

Let $\eta\in\Con^{1}_{s}\wed\Con^{1}_{(s,0)}\wed\Hor^{n,}{_{s+1}^{h}}\sub
\Var^{n+2}_{s+1}$,
then there is a unique morphism
\bEq\label{kappa}
K_{\eta} \in 
\Con^{1}_{(2s,s)}\otimes\Con^{1}_{(2s,0)}\wed\Hor^{n,}{_{2s+1}^{h}} 
\,,
\eEq
such that, for all $\Xi:\bY\to V\bY$,
$
E_{{j_{s}\Xi}\rfloor \eta} = C^{1}_{1} (j_{2s}\Xi\ten K_{\eta})$,
where $C^1_1$ stands for tensor
contraction on the first factor and $\rfloor$ denotes inner product 
(see \cite{KoVi03,Vit98}).
Furthermore, there is a unique pair of sheaf morphisms
\bEq\label{second}
H_{\eta} \in
\Con^{1}_{(2s,s)}\wed\Con^{1}_{(2s,0)}\wed\Hor^{n,}{_{2s+1}^{h}} \,,
\quad
G_{\eta} \in \Con^{2}_{(2s,s)}\wed\Hor^{n,}{_{2s+1}^{h}} \,,
\eEq
such that
${(\pi^{2s+1}_{s+1})}^*\eta = H_{\eta} - G_{\eta}$ and $H_{\eta}
= \frac{1}{2} \, A(K_{\eta})$,
where $A$ stands for antisymmetrisation.
Moreover, $G_{\eta}$ is {\em locally} of the type $G_{\eta} = d_H q_{\eta}$,
where
$q_{\eta} \in \Con^{2}_{(2s-1,s-1)}\wed\Hor^{n-1}{_{2s}}$, hence
$[\eta]=[H_{\eta}]$ \cite{KoVi03,Vit98}.

\bDf
Let $\gam \in \For^{n+1}_{s}$.
The morphism $H_{hd\gam}\equiv H_{[\cE_{n+1}(\gam)]}$, where square 
brackets denote equivalence class, is called the {\em
generalized Helmholtz  morphism}.
\END
\eDf

The standard Lie derivative of fibered morphisms with respect to a 
projectable vector field $j_{s}\Xi$ passes to the quotient in the 
variational sequence, thus defining a new quotient operator 
(introduced in \cite{FPV98a}), the {\em variational Lie derivative} 
$\cL_{j_{s}\Xi}$, acting on equivalence classes of fibered morphisms 
which are sections of the quotient sheaves in the variational 
sequence. Thus variational Lie derivatives of generalized Lagrangians 
or Euler--Lagrange morphisms can be conveniently represented as 
equivalence classes in $\Var^{n}_{s}$ and $\Var^{n+1}_{s}$. In 
particular, the following two results hold true \cite{FPV98a}.

\bTh\label{noether I}
Let $[\alp] = h(\alp)$ $\in$ $\Var^{n}_{s}$. Then we
have {\em locally} (up to pull-backs)
\beq
%\kap^{2s+1}_{s} \com 
\cL_{j_{s}\Xi}(h(\alp)) =
\Xi_{V} \rfloor \cE_{n}(h(\alp))+
d_{H}(j_{2s}\Xi_{V} \rfloor p_{d_{V}h(\alp)}+ \xi \rfloor h(\alp))\,.
\eeq
\eTh

\bTh\label{GeneralJacobi}
Let $\alp\in\For^{n+1}_{s}$. Then we have {\em globally} (up to pull-backs)
\beq
%\kap^{2s+1}_{s} \com 
\cL_{j_{s}\Xi} [\alp] =
\cE_{n}({j_{s+1}\Xi_{V} \rfloor h(\alp)}) +
C^{1}_{1}(j_{s}\Xi_{V}\ten K_{hd\alp}) \,.
\eeq
\eTh

%------------------------------------------------------------------------------------------------
\subsection{Generalized {\em gauge-natural} Jacobi morphisms}
%------------------------------------------------------------------------------------------------

We recall some previous results concerning
the representation of {\em generalized gauge-natural Jacobi morphisms} in
variational sequences and their relation with the second variation of a
generalized gauge-natural invariant Lagrangian \cite{PaWi03}.

\bDf\label{var}
Let $\alp:J_{s}\bY\to
\owed{p}T^*J_{s}\bY$. Let $\psi^{k}_{t_{k}}$, with $1\leq k\leq i$, be the
flows generated by an $i$--tuple
$(\Xi_{1},\ldots,\Xi_{i})$ of (vertical, although actually it is enough that
they are projectable) vector fields on $\bY$ and let $\Gam_{i}$ be the $i$--th
{\em formal variation} generated by the $\Xi_{k}$'s (to which we shall
refer as variation vector fields)  and defined,
for each $\by\in\bY$, by $\Gam_{i}(t_{1},\ldots,t_{i})(\by)=
\psi^{i}_{t_{i}}\circ \ldots \circ \psi^{1}_{t_{1}}(\by)$.
We define the $i$--th formal variation of the morphism $\alp$ to
be
\bEq
\del^{i}\alp \byd \frac{\der^{i}}{\der t_{1}\ldots
\der t_{i}}\big |_{t_{1},\ldots,
t_{i}=0}(\alp\circ j_{s}\Gam_{i}(t_{1},\ldots,t_{i})(\by)) \,.\END
\eEq
\eDf

The following Lemma states the relation between the $i$--th formal variation
of a morphism and its iterated Lie derivative
\cite{FPV02,PaWi03}.

\bLm\label{LIE}
Let $\alp: J_{s}\bY\to
\owed{p}T^*J_{s}\bY$ and $L_{j_{s}\Xi_{k}}$ be the Lie derivative
operator acting on differential fibered morphism.

Let
$\Gam_{i}$ be the $i$--th formal variation generated by variation
vector fields $\Xi_{k}$, $1\leq k\leq i$ on $\bY$. Then we have
\bEq
\del^{i} \alp = L_{j_{s}\Xi_{1}} \ldots L_{j_{s}\Xi_{i}} \alp \,.
\eEq
\eLm

Let  $\alp\in (\Var^{n}_{s})_{\bY}$. The operator $\del^{i}$ passes 
to the quotient in the variational sequence. We shall call the 
quotient
operator the {\em
$i$--th variational vertical derivative}.  We have
$\del^{i}[\alp]\byd [\del^{i}\alp] =[L_{\Xi_{i}} \ldots
L_{\Xi_{1}}\alp]=\cL_{\Xi_{i}} \ldots
\cL_{\Xi_{1}}[\alp]$.

Let now variation vector fields be vertical parts of  gauge-natural lifts.
By resorting to the Second Noether Theorem, we have the following
characterization of the second variational vertical derivative of a 
generalized Lagrangian \cite{PaWi03} which in fact enable us to 
relate the second variation with the morphism $K_{\eta}$ defined by 
Eq. \eqref{kappa} (for $\eta =hd\del\lam$).

First of all we fix some preliminary properties of gauge-natural lifts.
\bLm
Let $j_{s}\hat{\Xi}$ be the $s$-jet prolongation of $\hat{\Xi}$
which is a vector field on $J_{s}\bY_{\zet}$. It turns out then that
$j_{s}\mathfrak{G}(\bar{\Xi})=\mathfrak{G}(j_{s}\bar{\Xi})$.
\eLm

\bPf
Owing to linearity properties of
the Lie derivative of sections of gauge-natural bundles and since
$j_{s}\hat{\Xi}_{V}=-\pounds_{j_{s}\bar{\Xi}}$, the statement
is a consequence of Proposition 15.5 in \cite{KMS93}. \QED\ePf

Let
$j_{s}{\hat{\Xi}}_{V}$ be the vertical part according to the 
splitting \eqref{jet connection}. We shall denote
by $j_{s}\bar{\Xi}_{V}$ the induced section of the vector bundle
$\cA^{(r+s,k+s)}$ following the Lemma above. The set of all sections 
of this kind defines a vector subbundle of $J_{s}\cA^{(r,k)}$ which, 
by an abuse of notation,  we
shall denote by
$V J_{s}\cA^{(r,k)}$.
Let $j_{s}{\hat{\Xi}}_{V}$ be variation vector fields and let 
$\del^{2}_{\mathfrak{G}}\lam$ be the variation of $\lam$ with respect 
to such variation vector fields.

\bPr\label{x}
Let $\lam\in (\Var^{n}_{s})_{\bY}$ and let 
$\mathfrak{G}(\bar{\Xi})_{V}$ be a variation vector
field. Let $\chi(\lam,\mathfrak{G}(\bar{\Xi})_{V})\byd
C^{1}_{1} (j_{2s}\hat{\Xi}\ten 
K_{hd\cL_{j_{2s}\bar{\Xi}_V}\lam})\equiv E_{j_{s}\hat{\Xi}\rfloor
hd\cL_{j_{2s+1}\bar{\Xi}_V}\lam}$; then we have
\bEq
\del^{2}_{\mathfrak{G}}\lam = [\cE_{n}(j_{2s}\Xi \rfloor h\del\lam)
+\cJ(\lam,\mathfrak{G}(\bar{\Xi})_{V})] \,,
\eEq
where $[\, ]$ denotes the equivalence class in the variational 
sequence, while \\ $\cJ(\lam,\mathfrak{G}(\bar{\Xi})_{V})$ $\byd$
$E_{\chi(\lam,\mathfrak{G}(\bar{\Xi})_{V})}$.
\ePr

\bDf
Let $\bar{\Xi}\in \cA^{(r,k)}$.
We call the morphism $\cJ(\lam,\mathfrak{G}(\bar{\Xi})_{V})$ the {\em 
gauge-natural generalized Jacobi
morphism} associated with the Lagrangian $\lam$ and the gauge-natural 
lift $\mathfrak{G}(\bar{\Xi})_{V}$.
\END\eDf 

The morphism $\cJ(\lam,\mathfrak{G}(\bar{\Xi})_{V})$ is a
{\em linear} morphism with respect to the projection
$J_{4s}\bY_{\zet}\ucar{\bX}VJ_{4s}\cA^{(r,k)}\to J_{4s}\bY_{\zet}$. 
Notice that, {\em seen as a section of
$(\Var^{n}_{s})_{\bY\ucar{\bX}V\bY}$}, the equivalence class
$[\cE_{n}({j_{s}\bar{\Xi}_{V}
\rfloor  h(\del\lam)})]$ vanishes being a
local divergence of higher contact forms. This can also be compared 
with analogous results in
\cite{FPV02}.
Thus,
as a consequence Theorem \ref{GeneralJacobi} and Proposition \ref{x}, 
we have the following.

\bPr\label{comparison}
Let $\del^{2}_{\mathfrak{G}}\lam$ be the variation of $\lam$ with 
respect to vertical parts
of gauge-natural lifts of infinitesimal principal automorphisms. We have:
\bEq
\mathfrak{G}(\bar{\Xi})_{V}\rfloor
\cE_{n}(\mathfrak{G}(\bar{\Xi})_{V}\rfloor\cE_{n}(\lam))
=
\del^{2}_{\mathfrak{G}}\lam
=
\cE_{n}(\mathfrak{G}(\bar{\Xi})_{V}\rfloor
h(d\del\lam))\,.
\eEq
\ePr

%-----------------------------------------------------------------------------%
\section{Canonical covariant conserved currents}
%-----------------------------------------------------------------------------%

In the following we assume that the field equations are generated by
means of a variational principle from a Lagrangian which is
gauge-natural invariant, \ie invariant with respect to any
gauge-natural lift of infinitesimal right invariant vector fields.

\bDf\label{gn}
Let $(\hat{\Xi},\xi)$ be a projectable vector field on $\bY_{\zet}$.
Let $\lam \in \Var^{n}_{s}$
be a generalized Lagrangian. We say $\hat{\Xi}$ to be a {\em symmetry\/}
of $\lam$ if $\cL_{j_{s+1}\hat{\Xi}}\,\lam = 0$.

We say $\lam$ to be a
{\em gauge-natural invariant Lagrangian} if the gauge-natural lift
$(\hat{\Xi},\xi)$ of {\em any} vector
field $\bar{\Xi} \in \cA^{(r,k)}$ is a  symmetry for
$\lam$, \ie if $\cL_{j_{s+1}\bar{\Xi}}\,\lam = 0$.
In this case the projectable vector field
$\hat{\Xi}\equiv \mathfrak{G}(\bar{\Xi})$ is
called a {\em gauge-natural symmetry} of $\lam$.\END
\eDf

\bRm\label{fundRem}
As well known, the Second Noether Theorem deals 
with invariance properties of the Euler-Lagrange equations (so-called 
generalized symmetries or also Bessel-Hagen symmetries, see \eg the 
fundamental papers \cite{Tra67}). Although symmetries of a Lagrangian 
turn out to be also symmetries of the Euler--Lagrange morphism the 
converse is not true, in general.

In particular, although for a 
gauge-natural invariant Lagrangian $\lam$ we always have 
$\cL_{j_{s}\bar{\Xi}}\lam=0$,   $\cL_{j_{s}\bar{\Xi}_V}\lam$  does 
not need to be zero in principle; however when the second variation 
$\del^{2}_{\mathfrak{G}}\lam$ is required to vanish then 
$\cL_{j_{s}\bar{\Xi}_V} \cE_{n}(\lam)$ surely vanishes, \ie 
$j_{s}\bar{\Xi}_V$ is a generalized or Bessel--Hagen symmetry. The 
symmetries of the Euler--Lagrange morphism (Second Noether Theorem) 
impose some constraints on the conserved quantities associated with 
gauge-natural symmetries of $\lam$ (see \eg \cite{AnBe51}. 

\END\eRm

The First Noether Theorem takes a particularly interesting
form in the case of gauge-natural Lagrangians as shown in the following.

\bPr
\label{symmetry of L}
Let $\lam \in \Var^{n}_{s}$ be a gauge-natural Lagrangian and
$(\hat{\Xi},\xi)$
a gauge-natural symmetry of $\lam$. Then we have
$
0 = - \pounds_{\bar{\Xi}} \rfloor \cE_{n}(\lam)
+d_{H}(-j_{s}\pounds_{\bar{\Xi}}
\rfloor p_{d_{V}\lam}+ \xi \rfloor \lam) $.
Suppose that
$(j_{2s+1}\sig)^{*}(- \pounds_{\bar{\Xi}} \rfloor \cE_{n}(\lam)) = 0$.
Then, the $(n-1)$--form
$\eps = - j_{s}\pounds_{\bar{\Xi}} \rfloor p_{d_{V}\lam}+ \xi \rfloor \lam$
fulfills the equation $d ((j_{2s}\sig)^{*}(\eps)) = 0$.
\ePr

If $\sig$ is a critical section for $\cE_{n}(\lam)$, \ie
$(j_{2s+1}\sig)^{*}\cE_{n}(\lam) = 0$, the above equation
admits a physical interpretation as a so-called {\em weak conservation law}
for the density associated with $\eps$.

\bDf
Let $\lam \in \Var^{n}_{s}$ be a gauge-natural Lagrangian and
$\bar{\Xi} \in \cA^{(r,k)}$. Then the sheaf morphism $\eps: 
J_{2s}\bY_{\zet} \ucar{\bX} VJ_{2s}\cA^{(r,k)} \to 
\cC_{2s}^{*}[\cA^{(r,k)}]\ten\cC_{0}^{*}[\cA^{(r,k)}]
\wed (\owed{n-1} T^{*}\bX)$
is said to be a {\em gauge-natural weakly conserved current\/}.\END
\eDf

\bRm\label{arbitrary1}
In general, this conserved current is not uniquely defined. In fact,
it depends on the choice of $p_{d_{V}\lam}$, which is not unique, in 
general  (see
\cite{Vit98} and references quoted therein).
\END\eRm

In gauge-natural Lagrangian theories it is a well known procedure
to perform suitable integrations by
parts to decompose the conserved current $\eps$ into the sum of
a conserved current vanishing along solutions of the Euler--Lagrange equations,
the so--called {\em reduced current}, and the
formal divergence of a skew--symmetric (tensor) density called a {\em
superpotential} (which is defined modulo a divergence).
Within such a procedure, the generalized Bianchi identities
are in fact necessary and (locally) sufficient conditions for the 
conserved current
$\epsilon$ to be not only closed but also the divergence of a 
skew-symmetric (tensor) density
along solutions of the  Euler--Lagrange equations.

The following 
Lemma is a geometric version of the integration by parts procedure 
quoted above and it is based on a global decomposition formula of 
vertical morphisms due to Kol\'a\v{r} \cite{Kol83}.

\bLm\label{kol}
Let $\ome(\lam,\mathfrak{G}(\bar{\Xi})_{V})\byd \pounds_{\bar{\Xi}} 
\rfloor \cE_{n} (\lam): J_{2s}\bY_{\zet} \ucar{\bX} V 
J_{2s}\cA^{(r,k)}
\to 
\Con_{2s}^{*}[\cA^{(r,k)}]\ten\Con_{2s}^{*}[\cA^{(r,k)}]\ten\Con_{0}^{*}[\cA^{(r,k)}]\wed
(\owed{n}T^{*}\bX)$. Then we have {\em globally}
\beq
(\pi^{4s+1}_{s+1})^{*}\ome(\lam,\mathfrak{G}(\bar{\Xi})_{V}) = 
\bet(\lam,\mathfrak{G}(\bar{\Xi})_{V}) +
F_{\ome(\lam,\mathfrak{G}(\bar{\Xi})_{V})}\,,
\eeq
where
$
\bet(\lam,\mathfrak{G}(\bar{\Xi})_{V})
\equiv
E_{\ome(\lam,\mathfrak{G}(\bar{\Xi})_{V})}:$ \\
$: J_{4s}\bY_{\zet} \ucar{\bX} VJ_{4s}\cA^{(r,k)}
\to\Con_{2s}^{*}[\cA^{(r,k)}]\ten
\Con_{0}^{*}[\cA^{(r,k)}]\ten\Con_{0}^{*}[\cA^{(r,k)}]\wed
(\owed{n}T^{*}\bX) $
and {\em locally}, $F_{\ome(\lam,\mathfrak{G}(\bar{\Xi})_{V})} = 
D_{H}M_{\ome(\lam,\mathfrak{G}(\bar{\Xi})_{V})}$,
with
$
M_{\ome(\lam,\mathfrak{G}(\bar{\Xi})_{V})}: $\\
$: J_{4s-1}\bY_{\zet} \ucar{\bX} VJ_{4s-1}\cA^{(r,k)}) \to
\Con_{2s}^{*}[\cA^{(r,k)}]\ten
\Con_{2s-1}^{*}[\cA^{(r,k)}]\ten\Con_{0}^{*}[\cA^{(r,k)}]\wed
(\owed{n-1}T^{*}\bX)$.
\eLm
Coordinate expressions for the morphisms 
$\bet(\lam,\mathfrak{G}(\bar{\Xi})_{V}) $ and 
$M_{\ome(\lam,\mathfrak{G}(\bar{\Xi})_{V})}$ can be found by a 
backwards procedure (see \eg \cite{Kol83}). In particular, 
$\bet(\lam,\mathfrak{G}(\bar{\Xi})_{V})$ is nothing but the 
Euler--Lagrange morphism associated with the {\em new} Lagrangian 
$\ome(\lam,\mathfrak{G}(\bar{\Xi})_{V})$ defined on the fibered 
manifold $J_{2s}\bY_{\zet} \ucar{\bX} VJ_{2s}\cA^{(r,k)}\to \bX$.
In particular, we get the following {\em local} decomposition of 
$\ome(\lam,\mathfrak{G}(\bar{\Xi})_{V})$:
\bEq
\ome(\lam,\mathfrak{G}(\bar{\Xi})_{V}) = 
\bet(\lam,\mathfrak{G}(\bar{\Xi})_{V}) +
D_{H}\tilde{\eps}(\lam,\mathfrak{G}(\bar{\Xi})_{V}) \,,
\eEq
where we put $\tilde{\eps}(\lam,\mathfrak{G}(\bar{\Xi})_{V}) 
\equiv M_{\ome(\lam,\mathfrak{G}(\bar{\Xi})_{V})}$.

\bDf
We call the global morphism $\bet(\lam,\mathfrak{G}(\bar{\Xi})_{V}) 
\byd E_{\ome(\lam,\mathfrak{G}(\bar{\Xi})_{V})}$
the {\em generalized} {\em Bianchi morphism} associated
with the Lagrangian $\lam$.
\END\eDf

\bRm
For any $(\bar{\Xi},\xi)\in \cA^{(r,k)}$, as a consequence of the 
gauge-natural invariance of the Lagrangian, by
the Noether's First Theorem,  the morphism 
$\bet(\lam,\mathfrak{G}(\bar{\Xi})_{V}) \equiv
\cE_{n}(\ome(\lam,\mathfrak{G}(\bar{\Xi})_{V}))$ is {\em locally} 
identically vanishing.
We stress that these are just {\em local generalized Bianchi
identities}.
In particular, we have {\em locally} 
$\ome(\lam,\mathfrak{G}(\bar{\Xi})_{V})$ $=$
  $D_{H}\tilde{\eps}(\lam,\mathfrak{G}(\bar{\Xi})_{V})$
\cite{Vari1}.\END
\eRm
The form $\tilde{\eps}(\lam,\mathfrak{G}(\bar{\Xi})_{V})\equiv 
M_{\ome(\lam,\mathfrak{G}(\bar{\Xi})_{V})}$ is
called a {\em local reduced current}. It vanishes along any critical 
section.

The problem of the general covariance of generalized 
Bianchi identities for field theories was posed by Anderson and 
Bergman already in $1951$ (see \cite{AnBe51}).
Let now $\mathfrak{K} \byd 
\textstyle{Ker}_{\cJ(\lam,\mathfrak{G}(\bar{\Xi})_{V})}$
be the {\em kernel} of
the generalized gauge-natural morphism $\cJ(\lam,\mathfrak{G}(\bar{\Xi})_{V})$.
As a consequence of Proposition \ref{comparison} and of considerations
above, we have the following covariant characterization of the kernel 
of generalized Bianchi morphism, the detailed proof of which will 
appear in \cite{PaWi03}.

\bTh
The generalized Bianchi morphism is globally vanishing if and only
if $\del^{2}_{\mathfrak{G}}\lam\equiv\cJ(\lam,\mathfrak{G}(\bar{\Xi})_{V})=
0$, \ie if and only if
$\mathfrak{G}(\bar{\Xi})_{V}\in\mathfrak{K}$.
\eTh

The gauge-natural invariance of the
variational principle {\em in its whole}  enables us to solve the 
{\em intrinsic indeterminacy}  in the conserved charges associated 
with gauge-natural symmetries of Lagrangian field theories (in 
\cite{Mat03}, for example, the special case of the gravitational 
field coupled with fermionic matter is considered and the Kosmann 
lift is then invoked as an {\em ad hoc}  choice to recover the well 
known expression of the Komar superpotential). By requiring the 
second variation to vanish, \ie on the kernel of the Jacobi 
morphism, we express gauge-natural lits of infinitesimal principal 
automorphism in terms of
the corresponding infinitesimal diffeomorphisms (their projections) 
on the basis manifolds (see Theorem \ref{fundtheorem} below). This is 
well known to be of great  importance within the theory of Lie 
derivative of
sections of a gauge-natural bundle and notably for the Lie derivative 
of spinors (see \eg the review given in \cite{Mat03}).

\bTh\label{fundtheorem}
Let $\lam\in \Var^{n}_{s}$ be a gauge-natural invariant generalized 
Lagrangian and let
$\mathfrak{G}(\bar{\Xi})$ be a gauge-natural lift of the principal 
infinitesimal automorphism
$\bar{\Xi}\in\cA^{r,k}$, \ie a gauge-natural symmetry of $\lam$. Then 
$\bar{\Xi}\in\cA^{r,k}$ is related to its
projection $\xi\in\cT_{\bX}$ by the condition
\beq
(-1)^{|\bsig|}D_{\bsig}\,\left(D_{\bmu}
\bar{\Xi}^{j}_{V}\left(\der_{j}(\der^{\bmu}_{i}\lam) - \sum_{|\balp |
= 0}^{s-|\bmu |}
(-1)^{|\bmu +\balp |} \frac{(\bmu +
\balp)!}{\bmu ! \balp !}
D_{\balp}\der^{\balp}_{j}(\der^{\bmu}_{i}\lam)\right)\right)=0\,.
\eeq
\eTh

\bPf
We recall that given a vector field
$j_{s}\hat{\Xi}: J_{s}\bY_{\zet} \to TJ_{r}\bY_{\zet}$, the splitting
\eqref{jet connection} yields $j_{s}\hat{\Xi} \, \com \, \pi^{s+1}_{s} =
j_{s}\hat{\Xi}_{H} + j_{s}\hat{\Xi}_{V}$
where, if $j_{s}\hat{\Xi} = \hat{\Xi}^{\gam}\der_{\gam} + 
\hat{\Xi}^i_{\balp}\der^{\balp}_i$, then we
have $j_{s}\hat{\Xi}_{H} = \hat{\Xi}^{\gam}D_{\gam}$ and
$j_{s}\hat{\Xi}_{V} =
D_{\balp}(\hat{\Xi}^i - y^i_{\gam}\hat{\Xi}^{\gam}) 
\der^{\balp}_{i}$. Analogous considerations
hold true of course also for the unique corresponding invariant vector
field $j_{s}\bar{\Xi}$ on $W^{(r,k)}\bP$.
In particular, the condition $j_{s}\bar{\Xi}_{V} 
=D_{\balp}(\bar{\Xi}^{i}_{V})\der^{\balp}_{i}\in \mathfrak{K}$
implies, of course,  that the components $\bar{\Xi}^i_{\balp}$
and
$\bar{\Xi}^{\gam}$ {\em are not} independent, but they are {\em 
related} in such a way
that $j_{s}\bar{\Xi}_{V}=
D_{\balp}(\hat{\Xi}^i - y^i_{\gam}\hat{\Xi}^{\gam}) \der^{\balp}_{i}$ must be a
solution of generalized gauge-natural Jacobi equations for the Lagrangian
$\lam$.
\QED
\ePf

\bRm
For each $\bar{\Xi}\in \cA^{(r,k)}$ such that $\bar{\Xi}_{V}\in 
\mathfrak{K}$, we have
$\cL_{j_{s}\bar{\Xi}_{H}}\ome(\lam,
\mathfrak{K})=0$; the latter is a naturality condition for the 
morphism $\ome(\lam,
\mathfrak{K})$  and says something on the Hamiltonian structure of 
the theory itself (see \cite{FrPaWi04} for details).\END
\eRm

The result above reflects in the
theory of conserved currents and superpotentials for gauge-natural 
field theories, where the theory of Lie derivatives of sections of 
gauge-natural  bundles finds one of its main application.
In the following we shall refer to {\em canonical} globally defined 
objects (such as currents or corresponding
superpotentials) by their explicit dependence on $\mathfrak{K}$.

\bCr
Let $\lam \in \Var^{n}_{s}$ be a gauge-natural Lagrangian and
$j_{s}\hat{\Xi}_{V}\in \mathfrak{K}$
a gauge-natural symmetry of $\lam$.
Being $\bet(\lam, \mathfrak{K})\equiv 0$, we have, {\em globally}, 
$\ome(\lam,\mathfrak{K}) = D_{H}\eps(\lam, \mathfrak{K})$,
then the following holds:
\bEq\label{strong conservation}
D_{H}(\eps(\lam, \mathfrak{K})-\tilde{\eps}(\lam, \mathfrak{K}) = 0\,.
\eEq
\eCr

Eq. \eqref{strong conservation} is referred as a gauge-natural
`strong' conservation law for the {\em global} density $\eps(\lam, 
\mathfrak{K}) -\tilde{\eps}(\lam,
\mathfrak{K})$.

We can now state the following fundamental result about the existence and
{\em globality} of canonical gauge-natural superpotentials in the framework
of variational sequences.

\bTh\label{global}
Let $\lam \in \Var^{n}_{s}$ be a gauge-natural Lagrangian and
$(j_{s}\hat{\Xi},\xi)$ a gauge-natural symmetry of $\lam$. Then there exists
a global sheaf morphism
$\nu(\lam, \mathfrak{K})$ $\in$ 
$\left(\Var^{n-2}_{2s-1}\right)_{\bY_{\zet} \ucar{\bX}
\mathfrak{K}}$
such that
\beq
D_{H}\nu(\lam, \mathfrak{K}) = \eps(\lam, \mathfrak{K}) -\tilde{\eps}(\lam,
\mathfrak{K})\,.
\eeq
\eTh

\bDf
We define the sheaf morphism $\nu(\lam, \mathfrak{K})$ to be a
{\em canonical gauge-natural
superpotential} associated with $\lam$.
\END\eDf

\bEx ({\bf Einstein-Yang--Mills theory})
Let $\lam\in\Var^{n}_s$.
It is known (see {\em e.g.} \cite{Kol83}) that
\beq
& d_{V}\lam = (d_{V}\lam)^{\balp}_{i}\vartht^{i}_{\balp}\wed \ome \,,
\,
E_{d_{V}\lam}
= \cE(\lam)_{i}\vartht^{i}\wed \ome \,,
\,
p_{d_{V}\lam}
= p(\lam)^{\balp\mu}_{i}\vartht^{i}_{\balp}\wed \ome_{\mu}\,,
\\
& p(\lam)^{\bbet\mu}_{i}
= (d_{V}\lam)^{\balp}_{i} \qquad \bbet +\mu= \balp, |\balp|= s  \,,
\\
& p(\lam)^{\bbet\mu}_{i}
= (d_{V}\lam)^{\balp}_{i} - D_{\nu}p(\lam)^{\balp \nu}_{i}
\qquad \bbet+\mu=\balp, |\balp|= s-1  \,,
\\
& \cE(\lam)^{\balp}_{i}
= (d_{V}\lam)^{\balp}_{i}- D_{\nu}p(\lam)^{\balp\nu}_{i}\qquad |\balp|=0 \,.
\eeq
Furthermore,
$\cE(\lam)_{i}=\sum_{|\balp|\leq s}(-1)^{|\balp|}
D_{\balp}(d_{V}\lam)^{\balp}_{i}$.

Let $(\bP,\bX,\pi;\bG)$ be a principal bundle, $g$ a metric on
$\bX$, $\mathfrak{k}$ an $ad$--invariant metric on $\bG$.
Let $\ome$ be a principal connection and $F$ its $\mathfrak{g}$--valued
curvature $2$--form.
Let us now take the {\em gauge--natural} bundle
$\bY = Lor(\bX)\ucar{\bX}\bC$, where $Lor(\bX)$ is the
bundle of Lorentzian metrics
over space--time $\bX$ and $\bC$ is the affine
bundle of principal connections $\ome$ over $\bP$. Local coordinates on $\bY$
are given by $x^{\mu}, g^{\mu\nu},\ome^{i}_{\mu}$.
Let us consider the {\em gauge--natural} Lagrangian $\lam$ defined on
the gauge--natural bundle $J_{2}Lor(\bX)\ucar{\bX}J_{1}\bC$:
\bEq\label{unified}
\lam=\lam_{H}(g^{\mu\nu},R_{\mu\nu})+\lam_{YM}(g^{\mu\nu},F^{i}_{\mu\nu})\,,
\eEq
where $\lam_{H}=-\frac{1}{2\kappa}\sqrt{g}g^{\alp\bet}R_{\alp\bet}$ is
the Einstein Lagrangian, $R_{\alp\bet}$ is the Ricci tensor of the
metric $g$ given by
$R_{\alp\bet}\byd
R^{\mu}_{\alp\mu\bet}=D_{\mu}\gam^{\mu}_{\alp\bet}-D_{\bet}\gam^{\mu}_{\alp\mu}+
\gam^{\mu}_{\nu\mu}\gam^{\nu}_{\alp\bet}-
\gam^{\mu}_{\nu\bet}\gam^{\nu}_{\alp\mu}$, with
$\gam^{\mu}_{\nu\bet}=\frac{1}{2}g^{\mu\alp}(D_{\nu}g_{\bet\alp}-
D_{\alp}g_{\nu\bet}+D_{\bet}g_{\alp\nu})$ the Levi--Civita connection
of $g$, $\sqrt{g}=\sqrt{|det(g^{\mu\nu})|}$, $\kappa$ is a constant and
$\lam_{YM}(g_{\mu\nu},F^{i}_{\mu\nu})=
-\frac{1}{4}\sqrt{g}F^{\lam\gam}_{i}F_{\lam\gam}^{i}$ is the 
Yang--Mills Lagrangian. Here
$F^{\lam\gam}_{i}=\mathfrak{k}_{ij}g^{\lam\alp}g^{\gam\bet}F^{j}_{\alp\bet}$.

Notice that in this case $\bY_{\zet}= J_{2}Lor(\bX)\ucar{\bX}J_{1}\bC$
and the order of the gauge-natural bundle is $(r,k)=(3,2)$.
An infinitesimal right-invariant automorphism of $\bP$ is given
in the split form:
\beq
\Xi = \Xi_{h} + \Xi_{v}= \Xi^{\mu}(\der_{\mu} +
\ome^{i}_{\mu}\der_{i}) + (\Xi^{i} -
\ome^{i}_{\mu}\Xi^{\mu})\der_{i} \,,
\eeq
where $\der_{i}$ is a local basis of right-invariant vertical 
vector fields on $\bP$.
We shall respectively indicate by $\Xi_{v}$ and $\Xi_{h}$
the vertical and horizontal components of $\Xi$ with
respect to the principal connection $\ome$; we shall write
$\Xi^{i}_{v} = \Xi^{i} - \ome^{i}_{\mu}\Xi^{\mu}$. Notice that 
$\Xi^{i}_{v}$  is {\em not} equal to $\Xi^{i}_{V}$, but they are 
related in a clear and simple way.
We get $\eps^{\sig}(\lam,\Xi_{h})
=\eps^{\sig}(\lam_{H},\Xi_{h})+\eps^{\sig}(\lam_{YM},\Xi_{h})$,
where
\beq
\eps^{\sig}(\lam_{H},\Xi_{h})=
\frac{1}{\kappa}\sqrt{g}(R^{\sig}_{\bet}-Rg^{\sig}_{\bet})\Xi^{\bet}
+\nabla_{\mu}[\frac{\sqrt{g}}{2\kappa}
(\nabla^{\sig}\Xi^{\mu}-\nabla^{\mu}\Xi^{\sig})]\,.
\eeq
and
\beq
\eps^{\sig}(\lam_{YM},\Xi_{h})=
(2p^{\mu\sig}_{i}\pounds_{\Xi_{h}}\ome^{i}_{\mu}-\lam_{YM}\Xi^{\sig})=
-\sqrt{g}(F^{\mu\sig}_{i}F^{i}_{\mu\nu}-
\frac{1}{4}F^{\mu\rho}_{i}F^{i}_{\mu\rho}\del^{\sig}_{\nu})\Xi^{\nu}\,.
\eeq
$R$ is the scalar curvature and
$p^{\mu\nu}_{i}=-\frac{\sqrt{g}}{2}F^{\mu\nu}_{i}$.
Here and in the sequel $\nabla_{\mu}$ denotes the covariant
metric derivative with respect to $g$.

A ``horizontal'' superpotential is given by:
$\nu_{h}^{\sig\mu}=\frac{\sqrt{g}}{4\kappa}
(\nabla^{\sig}\Xi^{\mu}-\nabla^{\mu}\Xi^{\sig})$,
which is essentially the Komar superpotential \cite{Vari1}.

Furthermore, we have
$\eps^{\sig}(\lam_{YM},\Xi_{v})=
-2p^{\mu\sig}_{i}\nabla_{\mu}\Xi^{i}_{v}=
-\nabla_{\mu}(-2p^{\mu\sig}_{i}\Xi^{i}_{v})+
2\nabla_{\mu}p^{\mu\sig}_{i}\Xi^{i}_{v}$.
Then there exists a ``vertical'' superpotential, given by:
\beq
\nu_{v}^{\mu\sig}=p^{\mu\sig}_{i}\Xi^{i}_{v}=-\frac{\sqrt{g}}{2}F^{\mu\sig}_{i}
\Xi^{i}_{v}\,.
\eeq

From Theorem \ref{fundtheorem}, we deduce that also the components $\Xi^{i}_{v}$ 
-- when $\Xi_{V}$ is an element of the kernel of the 
gauge-natural generalized Jacobi morphism -- 
can be expressed invariantly in terms of the components 
$\Xi^{\mu}$ of the projection
of $\Xi_{v}$: 
\beq
(-1)^{|\bsig|}D_{\bsig}\,(D_{\bmu}
( \hat{\Xi}^{i}_{v} + (\hat{\ome}^{i}_{\gam}-
y^i_{\gam})\hat{\Xi}^{\gam})\psi_{ji}^{\bmu})=0\,,
\eeq
where $\psi_{ji}^{\bmu}=(\der_{j}(\der^{\bmu}_{i}\lam) - 
\sum_{|\balp | = 0}^{s-|\bmu |}(-1)^{|\bmu +\balp |} 
\frac{(\bmu + \balp)!}{\bmu !\balp !} 
D_{\balp}\der^{\balp}_{j}(\der^{\bmu}_{i}\lam))$ and $\hat{\ome}$ is  the gauge
natural prolongation of $\ome$ (see \cite{FFP01}). When such a dependence is expressed
explicitly, we write 
$\nu_{v}^{\mu\sig}=\nu_{v}^{\mu\sig}(\lam, \mathfrak{K})$.  

Equivalently, given
$\Xi^{i}_{v}$ as above, Theorem
\ref{fundtheorem} says us also that some constraint on the connection $\ome$ is 
there. It is maybe noteworthy that $\lam$ is the {\em total} Lagrangian
\eqref{unified}.
\eEx

{\em Acknowledgments.}
The authors wish to thank I. Kol\'a\v r and D. Krupka for
many interesting discussions and the unknown  referee for useful 
remarks which lead to improve the text.

%------------------------------------------------------------------------%

\end{document}